\documentclass[letterpaper,english]{article}

\usepackage[T1]{fontenc}
\usepackage[latin9]{inputenc}
\usepackage{amstext}
\usepackage{graphicx}
\usepackage{esint}

\makeatletter

\pdfpageheight\paperheight
\pdfpagewidth\paperwidth

\newcommand{\lyxaddress}[1]{
\par {\raggedright #1
\vspace{1.4em}
\noindent\par}
}

\date{}

\makeatother

\usepackage{babel}
\begin{document}

\title{Analog readout for optical reservoir computers}

\author{A. Smerieri$^{\text{1}}$, F. Duport$^{\text{1}}$, Y. Paquot$^{\text{1}}$,
B. Schrauwen$^{\text{2}}$, M. Haelterman$^{\text{1}}$, S. Massar$^{\text{3}}$ }

\maketitle

\lyxaddress{$^{\text{1}}$Service OPERA-photonique, Université Libre de Bruxelles
(U.L.B.), 50 Avenue F. D. Roosevelt, CP 194/5, B-1050 Bruxelles, Belgium}

\lyxaddress{$^{\text{2}}$Department of Electronics and Information Systems (ELIS),
Ghent University, Sint-Pietersnieuwstraat 41, 9000 Ghent, Belgium.}

\lyxaddress{$^{\text{3}}$Laboratoire d'Information Quantique (LIQ), Université
Libre de Bruxelles (U.L.B.), 50 Avenue F. D. Roosevelt, CP 225, B-1050
Bruxelles, Belgium}
\begin{abstract}
Reservoir computing is a new, powerful and flexible machine learning
technique that is easily implemented in hardware. Recently, by using
a time-multiplexed architecture, hardware reservoir computers have
reached performance comparable to digital implementations. Operating
speeds allowing for real time information operation have been reached
using optoelectronic systems. At present the main performance bottleneck
is the readout layer which uses slow, digital postprocessing. We have
designed an analog readout suitable for time-multiplexed optoelectronic
reservoir computers, capable of working in real time. The readout
has been built and tested experimentally on a standard benchmark task.
Its performance is better than non-reservoir methods, with ample room
for further improvement. The present work thereby overcomes one of
the major limitations for the future development of hardware reservoir
computers.
\end{abstract}

\section{Introduction}

The term ``reservoir computing'' encompasses a range of similar
machine learning techniques, independently introduced by H. Jaeger
\cite{Jaeger2001} and by W. Maass \cite{Maass2002}. While these
techniques differ in implementation details, they share the same core
idea: that one can leverage the dynamics of a recurrent nonlinear
network to perform computation on a time dependent signal without
having to train the network itself. This is done simply by adding
an external, generally linear readout layer and training it instead.
The result is a powerful system that can outperform other techniques
on a range of tasks (see for example the ones reported in \cite{Schrauwen07,Lukosevicius2009}),
and is significantly easier to train than recurrent neural networks.
Furthermore it can be quite easily implemented in hardware \cite{Fernando2003,Schurmann2004,Paquot2010},
although it is only recently that hardware implementations with performance
comparable to digital implementations have been reported \cite{Appeltant,Larger2012,Paquot2012}.

One great advantage of this technique is that it places almost no
requirements on the structure of the recurrent nonlinear network.
The topology of the network, as well as the characteristics of the
nonlinear nodes, are left to the user. The only requirements are that
the network should be of sufficiently high dimensionality, and that
it should have suitable rich dynamics. The last requirement essentially
means that the dynamics allows the exploration of a large number of
network states when new inputs come in, while at the same time retaining
for a finite time information on the previous inputs\emph{ }\cite{Legenstein2005}.
For this reason, the reservoir computers appearing in literature use
widely different nonlinear units, see for example \cite{Jaeger2001,Maass2002,Fernando2003,Vandoorne2010}
and in particular the time multiplexing architecture proposed in \cite{Paquot2010,Appeltant,Larger2012,Paquot2012}.

Optical reservoir computers are particularly promising, as they can
provide an alternative path to optical computing. They could leverage
the inherent high speeds and parallelism granted by optics, without
the need for strong nonlinear interaction needed to mimic traditional
electronic components. Very recently, optoelectronic reservoir computers
have been demonstrated by different research teams \cite{Paquot2012,Larger2012},
conjugating good computational performances with the promise of very
high operating speeds. However, one major drawback in these experiments,
as well as all preceding ones, was the absence of readout mechanisms:
reservoir states were collected on a computer and post-processed digitally,
severely limiting the processing speeds obtained and hence the applicability.

An analog readout for experimental reservoirs would remove this major
bottleneck, as pointed out in \cite{Woods2012}. The modular characteristics
of reservoir computing imply that hardware reservoirs and readouts
can be optimized independently and in parallel. Moreover, an analog
readout opens the possibility of feeding back the output of the reservoir
into the reservoir itself, which in turn allows the use of different
training techniques \cite{Sussillo2009} and to apply reservoir computing
to new categories of tasks, such as pattern generation \cite{Jaeger2007,Jaeger2004}. 

In this paper we present a proposal for the readout mechanism for
opto-electronic reservoirs, using an optoelectronic intensity modulator.
The design that we propose will drastically cut down their operation
time, specially in the case of long input sequences. Our proposal
is suited to optoelectronic or all-optical reservoirs, but the concept
can be easily extended to any experimental time-multiplexed reservoir
computer. The mechanism has been tested experimentally using the experimental
reservoir reported in \cite{Paquot2012}, and compared to a digital
readout. Although the results are preliminary, they are promising:
while not as good as those reported in \cite{Paquot2012}, they are
however already better than non-reservoir methods for the same task
\cite{Jaeger2004}.

\section{Reservoir computing and time multiplexing\label{sec:Reservoir-computing-and}}

\subsection{Principles of Reservoir Computing}

The main component of a reservoir computer (RC) is a recurrent network
of nonlinear elements, usually called ``nodes'' or ``neurons''.
The system typically works in discrete time, and the state of each
node at each time step is a function of the input value at that time
step and of the states of neighboring nodes at the previous time step.
The network output is generated by a readout layer - a set of linear
nodes that provide a linear combination of the instantaneous node
states with fixed coefficients. 

The equation that describes the evolution of the reservoir computer
is
\begin{equation}
x_{i}(n)=f(\alpha m_{i}u(n)+\beta\sum_{j=1}^{N}w_{ij}x_{j}(n-1))\label{eq:GeneralReservoir}
\end{equation}
 where $x_{i}(n)$ is the state of the $i$-th node at discrete time
$n$, $N$ is the total number of nodes, $u(n)$ is the reservoir
input at time $n$, $m_{i}$ and $w_{ij}$ are the connection coefficients
that describe the network topology, $\alpha$ and $\beta$ are two
parameters that regulate the network's dynamics, and $f$ is a nonlinear
function. One generally tunes $\alpha$ and $\beta$ to have favorable
dynamics when the input to be treated is injected in the reservoir.
The network output $y(n)$ is then constructed using a set of readout
weights $W_{i}$ and a bias weight $W_{b}$, as

\begin{equation}
y(n)=\sum_{i=1}^{N}W_{i}x_{i}(n)+W_{b}\label{eq:GeneralOutput}
\end{equation}

Training a reservoir computer only involves the readout layer, and
consists in finding the best set of readout weights $W_{i}$ and bias
$W_{b}$ that minimize the error between the desired output and the
actual network output. Unlike conventional recurrent neural networks,
the strength of connections $m_{i}$ and $w_{ij}$ are left untouched.
As the output layer is made only of linear units, given the full set
of reservoir states $x_{i}(n)$ for all the time steps $n$, the training
procedure is a basic, regularized linear regression.

\subsection{Time multiplexing}

The number of nodes in a reservoir computer determines an upper limit
to the reservoir performance \cite{Verstraeten2010}; this can be
an obstacle when designing physical implementations of RCs, which
should contain a high number of interconnected nonlinear units. A
solution to this problem proposed in \cite{Paquot2010,Appeltant},
is time multiplexing: the $x_{i}(n)$ are computed one by one by a
single nonlinear element, which receives a combination of the input
$u(n)$ and a previous state $x_{j}(n-1)$. In addition an input mask
$m_{i}$ is applied to the input $u(n)$, to enrich the reservoir
dynamics. The value of $x_{i}(n)$ is then stored in a delay line
to be used at a later time step $n+1$. The interaction between different
neurons can be provided by either having a slow nonlinear element
which couples state $x_{i}$ to the previous states $x_{i-1},x_{i-2},...$
\cite{Appeltant}, or by using an instantaneous nonlinear element
and desynchronizing the input with respect to the delay line \cite{Paquot2012}.

\subsection{Hardware RC with digital readout}

\begin{figure}
\includegraphics[scale=0.4]{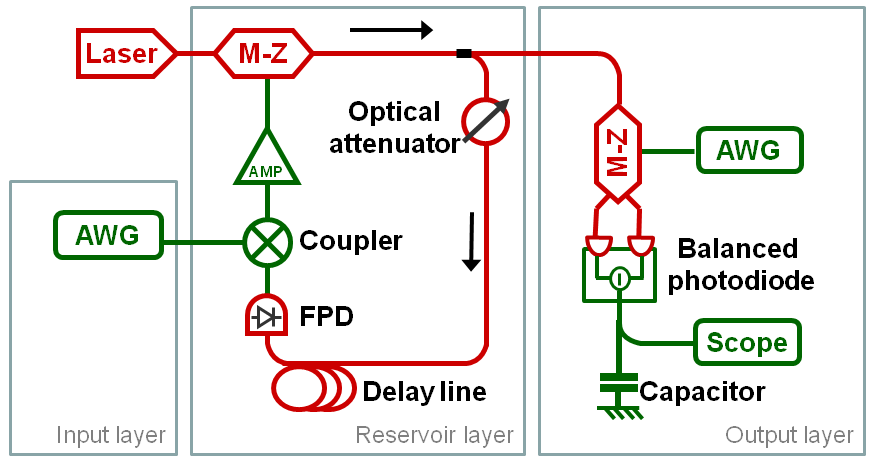}\caption{\label{fig:ExpScheme}Scheme of the experimental setup, including
the optoelectronic reservoir ('Input' and 'Reservoir' layers) and
the analog readout ('Output' layer). The red and green parts represent
respectively the optical and electronic components. ``AWG'': Arbitrary
waveform generator. ``M-Z'': $LiNbO_{3}$ Mach-Zehnder modulator.
``FPD'': Feedback photodiode. ``AMP'': Amplifier. ``Scope'':
NI PXI acquisition card.}
\end{figure}

The hardware reservoir computer we use in the present work is identical
to the one reported in \cite{Paquot2012} (see also \cite{Larger2012}).
It uses the time-multiplexing with desynchronisation technique described
in the previous paragraph. We give a brief description of the experimental
system, represented in the left part of Figure \ref{fig:ExpScheme}.
It uses a $LiNbO_{3}$ Mach-Zehnder (MZ) modulator, operating on a
constant power 1560 nm laser, as the nonlinear component. A MZ modulator
is a voltage controlled optoelectronic device; the amount of light
that it transmits is a sine function of the voltage applied to it.
The resulting state $x_{i}(n)$ is encoded in a light intensity level
at the MZ output. It is then stored in a spool of optical fiber, acting
as delay line of duration $T=8.5\mu s$, while all the subsequent
states $x_{i}(n)$ are being computed by the MZ modulator. When a
state $x_{i}(n)$ reaches the end of the fiber spool it is converted
into a voltage by a photodiode.

The input $u(n)\mbox{ }$is multiplied by the input mask $m_{i}$
and encoded in a voltage level by an Arbitrary Waveform Generator
(AWG). The two voltages corresponding to the state $x_{i}(n)$ at
the end of the fiber spool and the input $m_{i}u(n)$ are added, amplified,
and the resulting voltage is used to drive the MZ modulator, thereby
producing the state $x_{j}(n+1)$, and so on for all values of $n$.

In the experiment reported in \cite{Paquot2012} a portion of the
light coming out of the MZ is deviated to a second photodiode (not
shown in Figure \ref{fig:ExpScheme}), that converts it into a voltage
and sends it to a digital oscilloscope. The Mach-Zehnder output can
be represented as ``steps'' of light intensities of duration $\theta$
(see Figure \ref{fig:TypicalInput}a), each one representing the value
of a single node state $x_{i}$ at discrete time $n$. The value of
each $x_{i}(n)$ is recovered by taking an average of the measured
voltage for each state at each time step. The optimal readout weights
$W_{i}$ and bias $W_{b}$ are then calculated on a computer from
a subset (training set) of the recorded states, using ridge regression
\cite{Wyffels2008}, and the output $y(n)$ is then calculated using
equation \ref{eq:GeneralOutput} for all the states collected. The
performance of the reservoir is then calculated by comparing the reservoir
output $y(n)$ with the desired output $\hat{y}(n)$.

\section{Analog readout}

\subsection*{Readout scheme}

Developing an analog readout for the reservoir computer described
in section \ref{sec:Reservoir-computing-and} means designing a device
that multiplies the reservoir states shown in Figure \ref{fig:TypicalInput}a
by the readout weights $W_{i}$, and that sums them together in such
a way that the reservoir output $y(n)$ can be retrieved directly
from its output. However, this is not straightforward to do, since
obtaining good performance requires positive and negative readout
weights $W_{i}.$ In optical implementations \cite{Paquot2012,Larger2012}
the states $x_{i}$ are encoded as light intensities which are always
positive, so they cannot be subtracted one from another. Moreover,
the summation over the states must include only the values of $x_{i}$
pertaining to the same discrete time step $n$ and reject all other
values. This is difficult in time-multiplexed reservoirs, where the
states $x_{N}(n)$ and $x_{1}(n+1)$ follow seamlessly. 

Here we show how to resolve both difficulties using the scheme depicted
in the right panel of Figure \ref{fig:ExpScheme}. Reservoir states
encoded as light intensities in the optical reservoir computer and
represented in Figure \ref{fig:TypicalInput}a are fed to the input
of a second MZ modulator with two outputs. A second function generator
governs the bias of the second Mach-Zehnder, providing the modulation
voltage $V(t)$. The modulation voltage controls how much of the input
light passing through the readout Mach-Zehnder is sent to each output,
keeping constant the sum of the two output intensities. The two outputs
are connected to the two inputs of a balanced photodiode, which in
turn gives as output a voltage level proportional to the difference
of the light intensities received at its two inputs%
\footnote{A balanced photodiode consists of two photodiodes which convert the
two light intensities into two electric currents, followed by an electronic
circuit which produces as output a voltage proportional to the difference
of the two currents%
}. This allows us to multiply the reservoir states by both positive
and negative weights. 

The time average of the output voltage of the photodiode is obtained
by using a capacitor. The characteristic time of the analog integrator
$\tau$ is proportional to the capacity $C$.%
\footnote{In the case where the impedance of the coaxial cable $R=50\Omega$
is matched with the output impedance of the photodiode, we have $\tau=\frac{RC}{2}$%
} The role of this time scale is to include in the readout output all
the pertinent contributions and exclude the others. The final output
of the reservoir is the voltage across the capacitor at the end of
each discretized time $n$. 

What follows is a detailed description of the readout design.

\subsection*{Multiplication by arbitrary weights}

The multiplication of the reservoir states by arbitrary weights, positive
or negative, is realized by the second MZ modulator followed by the
balanced photodiode. The modulation voltage $V(t)$ that drives the
second Mach Zehnder is piecewise constant, with a step duration equal
to the duration $\theta$ of the reservoir states; transitions in
voltages and in reservoir states are synchronized. The modulation
voltage is also a periodic function of period $\theta N$, so that
each reservoir state $x_{i}(n)$ is paired with a voltage level $V_{i}$
that doesn't depend on $n$. The light intensities $O_{1}(t)$ and
$O_{2}(t)$ at the two outputs of the Mach-Zehnder modulator are
\begin{equation}
O_{1}(t)=I(t)\frac{1+\cos((V(t)+V_{bias})\frac{\pi}{V_{\pi}}+\varphi)}{2},O_{2}(t)=I(t)\frac{1-\cos((V(t)+V_{bias})\frac{\pi}{V_{\pi}}+\varphi)}{2},\label{eq:Output1MZ}
\end{equation}

where $I(t)$ is the light intensity coming from the reservoir, $V_{bias}$
is a constant voltage that drives the modulator, $\varphi$ is an
arbitrary, constant phase value, and $V_{\pi}$ is the half-wave voltage
of the modulator. Neglecting the effect of any bandpass filter in
the photodiode, and choosing $V_{bias}$ appropriately, the output
$P(t)$ from the photodiode can be written as 
\begin{equation}
P(t)=G(O_{1}(t)-O_{2}(t))=I(t)(G\sin(\frac{V(t)\pi}{V_{\pi}}))=I(t)W(t)\label{eq:OutputPhotodiode}
\end{equation}

with $G\mbox{ }$a constant gain factor. In other words, by setting
the right bias and driving the modulator with a voltage $V(t)$, we
multiply the signal $I(t)$ by an arbitrary coefficient $W(t)$. Note
that, if $V(t)$ is piecewise constant, then $W(t)$ is as well. This
allows us to achieve the multiplication of the states $x_{i}(n)$,
encoded in the light intensity $I(t)$, by the weights $W_{i}$, just
by choosing the right voltage $V(t)$, as shown in Figure \ref{fig:TypicalInput}b.

\subsection*{Summation of weighted states}

To achieve the summation over all the states pertaining to the same
discrete time step $n,$ which according to equation \ref{eq:GeneralOutput}
will give us the reservoir output minus the bias $W_{b}$, we use
the capacitor at the right side of the Output layer in Figure \ref{fig:ExpScheme}.
The capacitor provides the integration of the photodiode output given
by eq. \ref{eq:OutputPhotodiode} with an exponential kernel and time
constant $\tau$. If $\tau$ is significantly less than the amount
of time $\theta N$ needed for the system to process all the nodes
relative to a single time step, we can minimize the crosstalk between
node states relative to different time steps.

Let us consider the input $I(t)$ of the readout, and let $t=0$ be
the instant where the state of the first node for a given discrete
time step $n$ begins to be encoded in $I(t)$ . Using equation \ref{eq:OutputPhotodiode},
we can write the voltage $Q(t)$ on the capacitor at time $\theta N$
as
\begin{equation}
Q(\theta N)=Q(0)e^{-\frac{\theta N}{\tau}}+\int_{0}^{\theta N}I(s)W(s)e^{-\frac{\theta N-s}{\tau}}ds\label{eq:OutputIntegral}
\end{equation}
For $0<t<\theta N$, we have
\begin{equation}
I(t)=x_{i}(n),W(t)=w_{i},\:\mbox{for}\:\theta(i-1)<t<\theta i\label{eq:IinformofX}
\end{equation}

Integrating equation \ref{eq:OutputIntegral} yields
\begin{equation}
Q(\theta N)=Q(0)e^{-\frac{\theta N}{\tau}}+\sum_{i=1}^{N}x_{i}(n)\eta_{i}w_{i},\:\:\eta_{i}=e^{-\frac{\theta(N-i)}{\tau}}(1-e^{-\frac{\theta}{\tau}})\tau\label{eq:DiscretizedReadoutOutput}
\end{equation}

Equation \ref{eq:DiscretizedReadoutOutput} shows that, at time $\theta N$,
the voltage on the capacitor is a linear combination of the reservoir
states for the discrete time $n$, with node-dependent coefficients
$\eta_{i}w_{i}$, plus a residual of the voltage at time $0$, multiplied
by an extinction coefficient $e^{-\frac{\theta N}{\tau}}$. At time
$2\theta N$ the voltage on the capacitor would be a linear combination
of the states for discrete time $n+1$, multiplied by the same coefficients,
plus a residual of the voltage at time $\theta N$, and so on for
all values of $n$ and corresponding multiples of $\theta N$. 

\begin{figure}
\includegraphics[width=6in]{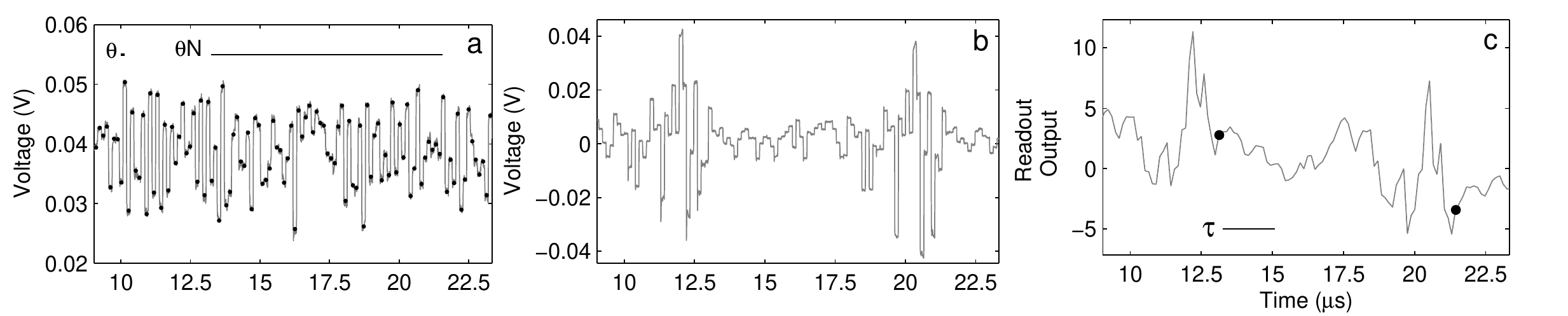}\caption{\label{fig:TypicalInput}a) Reservoir output $I(t)$. The gray line
represents the output as measured by a photodiode and an oscilloscope.
We indicated for reference the time $\theta=130ns$ used to process
a single node and the duration $\theta N=8.36\mu s$ of the whole
set of states. b) Output $P(t)$ of the balanced photodiode (see equation
\ref{eq:OutputPhotodiode}), with the trace of panel a) as input,
before integration. c) Voltage $Q(t)$ on the capacitor for the same
input (see equation \ref{eq:OutputIntegral}). The integration time
$\tau$ is indicated for reference. The black dots indicate the values
at the end of each discretized time $n$, taken as the output $y(n)$of
the analog readout.}
\end{figure}

A simple procedure would encode the weights $w_{i}=\frac{W_{i}}{\eta_{i}}$
onto the voltage $V(t)$ that drives the modulator , provide an external,
constant bias $W_{b}$, and have the output $y(n)$ of the reservoir,
defined by equation \ref{eq:GeneralOutput}, effectively encoded on
the capacitor. This simple procedure would however be unsatisfactory
because unavoidably some of the $\eta_{i}$ would be very small, and
therefore the $w_{i}$ would be large, spanning several orders of
magnitude. This is undesirable, as it requires a very precise control
of the modulation voltage $V(t)$ in order to recreate all the $w_{i}$
values, leaving the system vulnerable to noise and to any non-ideal
behavior of the modulator itself. 

To mitigate this, we adapt the training algorithm based on ridge regression
to our case. We redefine the reservoir states as $\xi_{i}(n)=x_{i}(n)\eta_{i}$;
we then calculate the weights $\omega_{i}$ that, applied to the states
$\xi_{i}$, give the best approximation to the desired output $\hat{y}(n)$.
The advantage here is that ridge regression keeps the norm of the
weight vector to a minimum; by redefining the states, we can take
the $\eta_{i}$ into account without having big values of $w_{i}$
that force us to be extremely precise in generating the readout weights. 

A sample trace of the voltage on the capacitor is shown in Figure
\ref{fig:TypicalInput}c.

\subsection*{Hardware implementation}

To implement the analog readout, we started from the experimental
architecture described in Section \ref{sec:Reservoir-computing-and},
and we added the components depicted in the right part of Figure \ref{fig:ExpScheme}.
For the weight multiplication, we used a second Mach-Zehnder modulator
(Photline model MXDO-LN-10 with bandwidth in excess of 10GHz and $V_{\pi}=5.9V$),
driven by a Tabor 2074 Arbitrary Waveform Generator (maximum sampling
rate 200 MSamples/s). The two outputs of the modulator were fed into
a balanced photodiode (Terahertz technologies model 527 InGaAs balanced
photodiode, bandwidth set to 125MHz, response set to 1000V/W), whose
output was read by the National Instruments PXI digital acquisition
card (sampling rate 200 MSamples/s).

In most of the experimental results described here, the capacitor
at the end of the circuit was simulated and not physically inserted
into the circuit: this allowed us to quickly cycle in our experiments
through different values of $\tau$ without taking apart the circuit
every time. The external bias $W_{b}$ to the output, introduced in
equation \ref{eq:GeneralOutput}, was also provided after the readout.
The reasoning behind these choices is that both these implementations
are straightforward, while the use of a modulator and a balanced photodiode
as a weight generator is more complex: we chose to focus on the latter
issue for now, as our goal is to validate the proposed architecture.

\section{Results}

As a benchmark for our analog readout, we use a wireless channel equalization
task, introduced in 1994 \cite{Mathews1994} to test adaptive bilinear
filtering and subsequently used by Jaeger \cite{Jaeger2004} to show
the capabilities of reservoir computing. This task is becoming a standard
benchmark task in the reservoir computing community, and has been
used for example in \cite{Rodan2011}. It consists in recovering a
sequence of symbols transmitted along a wireless channel, in presence
of multiple reflections, noise and nonlinear distortion; a more detailed
description of the task can be found in the Appendix. The performance
of the reservoir is usually measured in Symbol Error Rate (SER), i.e.
the rate of misinterpreted symbols, as a function of the amount of
noise in the wireless channel. 

\begin{figure}
\includegraphics[width=6in]{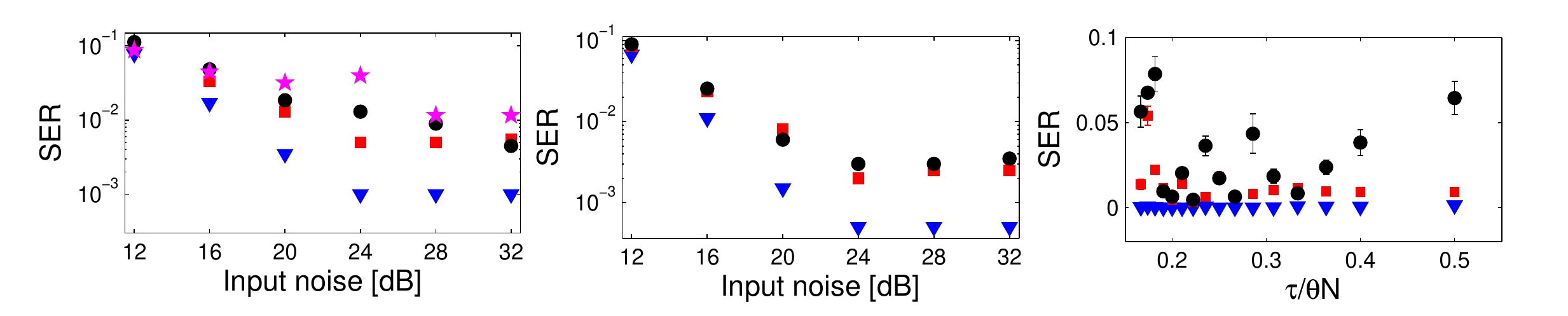}

\caption{\label{fig:results} Performance of the analog readout. Left: Performance
as a function of the input SNR, for a reservoir of 28 nodes, with
$\tau/\theta N=0.18$. Middle: Performance for the same task, for
a reservoir of 64 nodes, $\tau/\theta N=0.18$. Right: Performance
as a function of the ratio $\tau/\theta N$, at constant input noise
level (28 dB SNR) for a reservoir of 64 nodes. The performance is
measured in Signal Error Rate (SER). Blue triangles: reservoir with
digital readout. Red squares: reservoir with ideal analog readout.
Black circles: reservoir with experimental analog readout (simulated
capacitor). Purple stars in the left panel: reservoir where a physical
capacitor has been used.}
\end{figure}

Figure \ref{fig:results} shows the performance of the experimental
setup of \cite{Paquot2012} for a network of 28 nodes and one of 64
nodes, for different amounts of noise. For each noise level, three
quantities are presented. The first is the performance of the reservoir
with a digital readout (blue triangles), identical to the one used
in \cite{Paquot2012}. The second is the performance of a simulated,
ideal analog readout, which takes into account the effect of the $\eta_{i}$
coefficients introduced in equation \ref{eq:DiscretizedReadoutOutput},
but no other imperfection. It produces as output the discrete sum
$\omega_{b}+\sum_{i=1}^{N}\xi_{i}\omega_{i}$ (red squares). This
is, roughly speaking, the goal performance for our experimental readout.
The third and most important is the performance of the reservoir as
calculated on real data taken from the analog reservoir with the analog
output, with the effect of the continuous capacitive integration computed
in simulation (black circles). 

As can be seen from the figure, the performance of the analog readout
is fairly close to its ideal value, although it is significantly worse
than the performance of the digital readout. However, it is already
better than the non-reservoir methods reported in \cite{Mathews1994}
and used by Jaeger as benchmarks in \cite{Jaeger2004}. It can also
handle higher signal-to-noise ratios. As expected, networks with more
nodes have better performance; it should be noted, however, that in
experimental reservoirs the number of nodes cannot be raised over
a certain threshold. The reason is that the total loop time $\theta N$
is determined by the experimental hardware (specifically, the length
of the delay line); as $N$ increases, the length $\theta$ of each
node must decrease. This leaves the experiment vulnerable to noise
and bandpass effect, that may lead, for example, to an incorrect discretization
of the $x_{i}(n)$ values, and an overall worse performance.

We did test our readout with a $70nF$ capacitor, with a network of
28 nodes, to prove that the physical implementation of our concept
is feasible: the performance of this setup is shown in the left panel
of Figure \ref{fig:results}. The results are comparable to those
obtained in simulation, even if, at low levels of noise in the input,
the performance of the physical setup is slightly worse.

The rightmost panel of figure \ref{fig:results} shows the effects
of the choice of the capacitor at the end of the circuit, and therefore
of the value of $\tau$. The plot represents the performance at 28
dB SNR for a network of 64 nodes, for different values of the ratio
$\tau/\theta N$, obtained by averaging the results of 10 tests. It
is clear that the choice of $\tau$ has a complicated effect on the
readout performance; however, some general rules may be inferred.
Too small values of $\tau$ mean that the contribution from the very
first nodes is vanishingly small, effectively decreasing the reservoir
dimensionality, which has a strong impact on the performance both
of the ideal and the experimental reservoir. On the other hand, larger
values of $\tau$ impact the performance of the experimental readout,
as the residual term in equation \ref{eq:DiscretizedReadoutOutput}
gets larger. A compromise value of $\tau/\theta N=0.222$ seems to
give the best result, corresponding in our case to a capacity of about
$70$ nF.

\section{Discussion}

To our knowledge, the system presented here is the first analog readout
for an experimental reservoir computer. While the results presented
here are preliminary, and there is much optimization of experimental
parameters to be done, the system already outperforms non-reservoir
methods. We expect to extend easily this approach to different tasks,
already studied in \cite{Larger2012,Paquot2012}, including a spoken
digit recognition task on a standard dataset.%
\footnote{Texas Instruments-Developed 46-Word Speaker-Dependent Isolated Word
Corpus (TI46), September 1991, NIST Speech Disc 7-1.1 (1 disc) (1991).%
} 

Further performance improvements can reasonably be expected from fine-tuning
of the training parameters: for instance the amount of regularization
in the ridge regression procedure, that here is left constant at $1\cdot10^{-4}$,
should be tuned for best performance. Adaptive training algorithms,
such as the ones mentioned in \cite{Legenstein2010}, could also take
into account nonidealities in the readout components. Moreover the
choice of $\tau,$ as Figure \ref{fig:results} shows, is not obvious
and a more extensive investigation could lead to better performance. 

The architecture proposed here is simple and quite straightforward
to realize. It is very modular, meaning that it can be added at the
output of any preexisting time multiplexing reservoir with minimal
effort, whether it is based on optics or electronics. The capacitor
at the end of the circuit could probably be substituted with a more
complicated, active electronic circuit performing the summation of
the incoming signal before resetting itself. This would eliminate
the problem of residual voltages, and allow better performance at
the cost of increased complexity of the readout.

The main interest of the analog readout is that it allows optoelectronic
reservoir computers to fully leverage their main characteristic, which
is the speed of operation. Indeed, removing the need for slow, offline
postprocessing is indicated in \cite{Woods2012} as one of the major
challenges in the field. Once the training is finished, optoelectronic
reservoirs can process millions of nonlinear nodes per second \cite{Paquot2012};
however, in the case of a digital readout, the node states must be
recovered and postprocessed to obtain the reservoir outputs. It takes
around $1.6$ seconds for the digital readout in our setup to retrieve
and digitize the states generated by a 9000 symbol input sequence.
The analog readout removes the need for postprocessing, and can work
at a rate of about $8.5$ $\mu s$ per input symbol, five orders of
magnitude faster than the electronic reservoir reported in \cite{Appeltant}. 

Finally, having an analog readout opens the possibility of feedback
- using the output of the reservoir as input or part of an input for
the successive time steps. This opens the way for different tasks
to be performed \cite{Jaeger2007} or different training techniques
to be employed \cite{Sussillo2009}.

\section*{Acknowledgments}

This research was supported by the Interuniversity Attraction Poles
program of the Belgian Science Policy Office, under grant IAP P7-35
«photonics@be»

\section*{Appendix: Nonlinear Channel Equalization task}

What follows is a detailed description of the channel equalization
task used to test the reservoir performance. The goal for this task
is to reconstruct a sequence $d(n)$ of symbols taken from $\{-3,-1,1,3\}$.
The symbols in $d(n)$ are mixed together in a new sequence $q(n)$
given by
\begin{eqnarray}
q(n) & = & 0.08d(n+2)-0.12d(n+1)+d(n)+0.18d(n-1)-0.1d(n\text{-}2)\label{eq:mixingsequence}\\
 &  & +0.091d(n-3)\text{-}0.05d(n-4)+0.04d(n-5)+0.03d(n-6)+0.01d(n\text{-}7)\nonumber 
\end{eqnarray}
which models a wireless signal reaching a receiver through different
paths with different traveling times. A noisy, distorted version $u(n)$
of the mixed signal $q(n)$, simulating the nonlinearities and the
noise sources in the receiver, is created by having $u(n)=q(n)+0.036q(n)^{2}-0.011q(n)^{3}+\nu(n)$,
where $\nu(n)$ is an i.i.d. Gaussian noise with zero mean adjusted
in power to yield signal-to-noise ratios ranging from 12 to 32 dB.
The sequence $u(n)$ is then fed to the reservoir as an input; the
output of the readout $R(n)$ is rounded off to the closest value
among $\{-3,-1,1,3\}$, and then compared to the desired symbol $d(n)$.
The performance is usually measured in Signal Error Rate (SER), or
the rate of misinterpreted symbols.

\end{document}